\documentclass[pre,showpacs,showkeys,preprintnumbers,amsmath,amssymb,superscriptaddress,onecolumn]{revtex4-2}

\usepackage{graphicx}
\usepackage{bm}
\usepackage{color}
\usepackage{epstopdf}
\usepackage{amsmath} 
\usepackage{lastpage,fancyhdr,graphicx}
\usepackage[breaklinks=true,colorlinks=true,linkcolor=blue,urlcolor=blue,citecolor=blue]{hyperref}

\begin{document}

\title{Diffusion-controlled reactions: an overview}

\author{Denis~S.~Grebenkov}
 \email{denis.grebenkov@polytechnique.edu}
\affiliation{
Laboratoire de Physique de la Mati\`{e}re Condens\'{e}e, \\ 
CNRS -- Ecole Polytechnique, Institut Polytechnique de Paris, 91120 Palaiseau, France}

\date{\today}

\begin{abstract}
We review the milestones in the century-long development of the theory
of diffusion-controlled reactions.  Starting from the seminal work by
von Smoluchowski who recognized the importance of diffusion in
chemical reactions, we discuss perfect and imperfect surface
reactions, their microscopic origins, and the underlying mathematical
framework.  Single-molecule reaction schemes, anomalous bulk
diffusions, reversible binding/unbinding kinetics and many other
extensions are presented.  An alternative encounter-based approach to
diffusion-controlled reactions is introduced, with emphasis on its
advantages and potential applications.  Some open problems and future
perspectives are outlined.
\end{abstract}


\def\x{\bm{x}}
\def\erfc{\mathrm{erfc}}
\def\P{\mathbb{P}}

\newcommand{\clr}{\color{red}}
\newcommand{\clb}{\color{blue}}


\maketitle

\section{Introduction}

The nineteenth century was marked by impressive advances in the theory
of chemical reactions, even though the existence of atoms and
molecules, the (quantum) origins of chemical bonds and many other
fundamental aspects remained to be clarified.  Understanding of
stoichiometric relations between reactants and the development of a
mathematical theory of ordinary differential equations (ODE) provided
a powerful tool to describe the kinetics of very sophisticated
reactions.  On a basic level, stoichiometric relations allow one to
calculate the right proportions of ingredients and the masses of
produced reactants at the end.  Moreover, they determine the form of
the ODEs that govern the time evolution of concentrations of the
reactants \cite{House}.  For instance, upon disintegration of a
substance $A$, its concentration $[A]$ obeys the simplest ODE,
\begin{equation}  \label{eq:[A]-linear}
\frac{d[A]}{dt} = -k_A [A],
\end{equation}
where $k_A$ is the disintegration rate; here, change in time of the
concentration on the left-hand side is proportional to the remaining
concentration on the right-hand side.  The solution of this equation,
$[A](t) = [A]_0 \exp(- k_A t)$, shows an exponential decay of the
concentration from the initial level $[A]_0$.  The simplicity of this
solution is caused by {\it linearity} of the equation.  For instance,
the dynamics of a bimolecular synthesis reaction,
\begin{equation}   \label{eq:AB->AB}
A + B \rightarrow AB,
\end{equation}
is described by {\it nonlinear} differential equations such as
\begin{equation}
\frac{d[A]}{dt} = -k_{AB} [A][B],
\end{equation}
in which the rate of decrease of the concentration $[A]$ is
proportional to the product of concentrations of both substances,
i.e., to the likelihood of meeting between reactants $A$ and $B$.
More generally, stoichiometric relations, which determine how many
copies of each reactant molecule are involved in chemical reaction,
set the powers of the involved concentrations.
The nonlinearity of ODEs describing chemical kinetics presents one of
the major mathematical challenges for their analysis but also the
origin of many peculiar features (e.g., non-existence or
non-uniqueness of the solution, a finite time to the extinction of
some reactants, etc.).  These features and their implications in
chemistry and biology have been thoroughly investigated in the
twentieth century \cite{Murrey,Volpert}.

The above description totally ignores spatial aspects of chemical
reactions, as if the concentrations of reactants were homogeneous in
space at any time.  This is known as the {\it well-mixing assumption}
when the reactants are assumed to be well mixed so that reaction
occurs in different points of space in the same way.  However, there
are numerous situations, in which the spatial aspects are critically
important.  For instance, many biochemical reactions in living cells
involve proteins and macromolecules that are produced in one spatial
location but have to diffuse to another location to find their
reaction partners (e.g., receptors, enzymes, or specific sites on DNA
chains).  Even for small particles such as oxygen molecules, ions and
metabolites, there is generally a gradient of concentration between
their ``source'' and ``sink'' that drives their directional transport
in space.  Moreover, even if the concentrations $[A]$ and $[B]$ are
macroscopically homogeneous but low, single molecules $A$ and $B$ have
to meet each other to form an aggregate $AB$ according to the reaction
(\ref{eq:AB->AB}), and this transport step takes time and can be the
limiting factor in the overall reaction rate.  The crucial role of
diffusion was put forward by M. von Smoluchowski, who formulated in
1917 the first mathematical description of the coagulation dynamics
\cite{Smoluchowski18}, which later became the cornerstone of the
theory of diffusion-controlled reactions in a much broader context
\cite{North66,Wilemski73,Calef83,Berg85,Rice85,Lindenberg}.  
Examples of diffusion-controlled reactions include coagulation
dynamics \cite{Smoluchowski18,Witten81}, most catalysis and enzymatic
reactions \cite{Lauffenburger,Kuchler16} and ligand-protein
associations \cite{Hill75,Zwanzig90,Held11}, geminate recombination of
radicals and ions \cite{Sano79,Agmon88}, reactions in micellar and
vesicular systems \cite{Sano81}, spin relaxation on magnetic
impurities \cite{Brownstein79,Grebenkov07}, diffusive search by a
transcription factor protein for a specific binding site on a DNA
molecule \cite{Richter74,Berg81,Sheinman12}, control of flux by narrow
passages and hidden targets in cellular biology
\cite{Holcman13,Bressloff13}, self-propulsion of active colloids
\cite{Golestanian09,Oshanin17}, and oxygen capture in the lungs
\cite{Weibel,Sapoval02,Grebenkov05}.
Note that such reactions bear other names as diffusion-limited,
diffusion-mediated, diffusion-assisted, or diffusion-influenced
reactions.  In the past, these names were sometimes used to
distinguish the role of diffusion, e.g., whether the reaction occurs
instantly upon the first encounter of the reactants, or after
additional chemical kinetics step.  We do not make such distinctions
and understand diffusion-controlled reactions in a broad sense as
reactions in which diffusion is relevant.

In this concise review, we focus on the spatial aspect of chemical
reactions.  In Sec. \ref{sec:diffusion}, we describe a chemical
transformation on a catalytic surface and emphasize the role of
diffusion and the consequent spatial dependence of the concentration
(e.g., the formation of a depletion zone).  Section \ref{sec:Robin}
presents a more realistic setting of imperfect surface reactions,
which combine diffusion in the bulk and chemical kinetics on the
surface.  In Sec. \ref{sec:extension}, we briefly overview various
extensions such as anomalous diffusions, reversible binding/unbinding
reactions, reactions in dynamically heterogeneous media, etc.  Section
\ref{sec:encounter} describes an alternative approach to
diffusion-controlled reactions based on the statistics of encounters
between the reactant and the catalytic surface, while
Sec. \ref{sec:conclusion} concludes the review.

\section{The role of diffusive transport}
\label{sec:diffusion}

\begin{figure}[t!]
\includegraphics[width=0.19\textwidth]{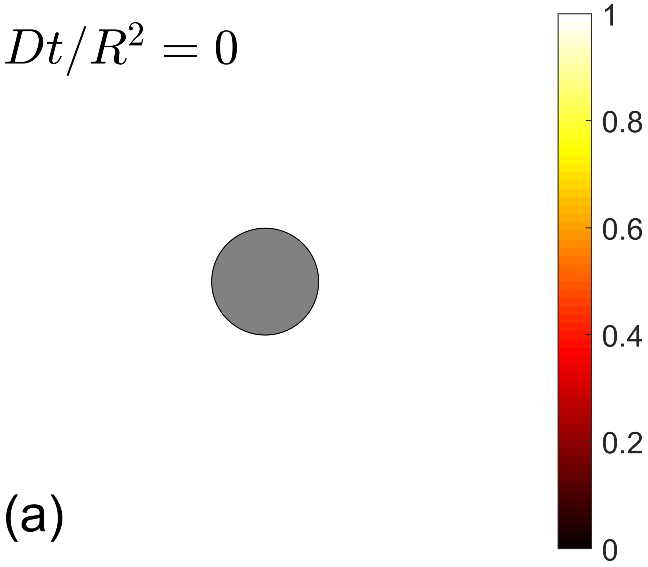}
\includegraphics[width=0.19\textwidth]{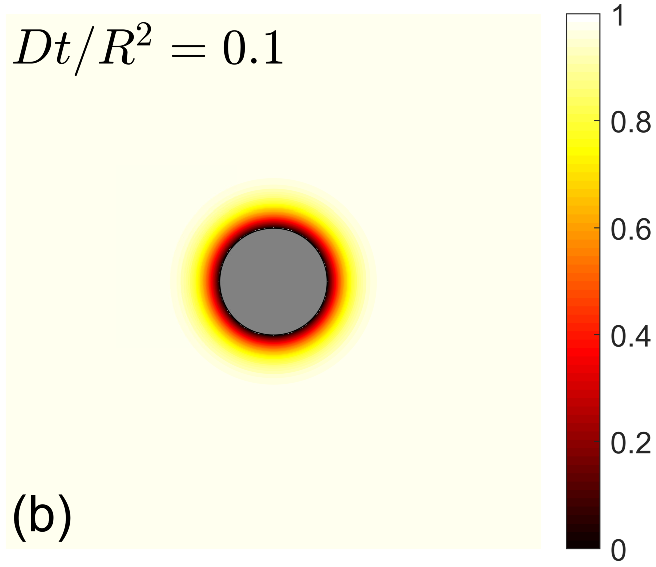}
\includegraphics[width=0.19\textwidth]{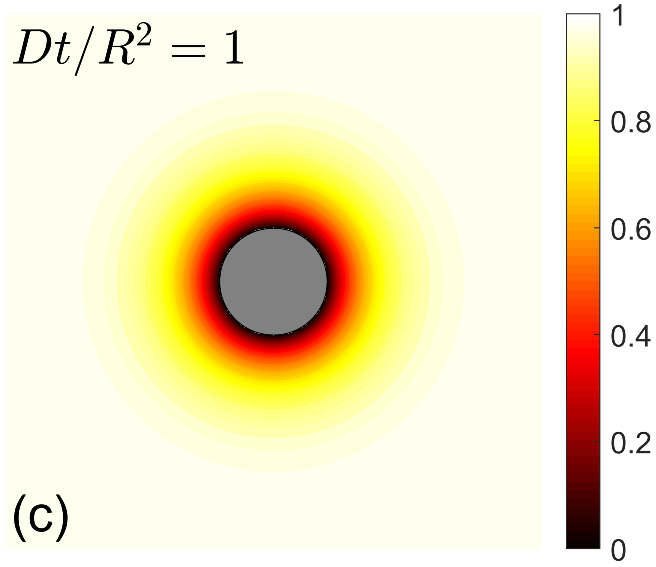}
\includegraphics[width=0.19\textwidth]{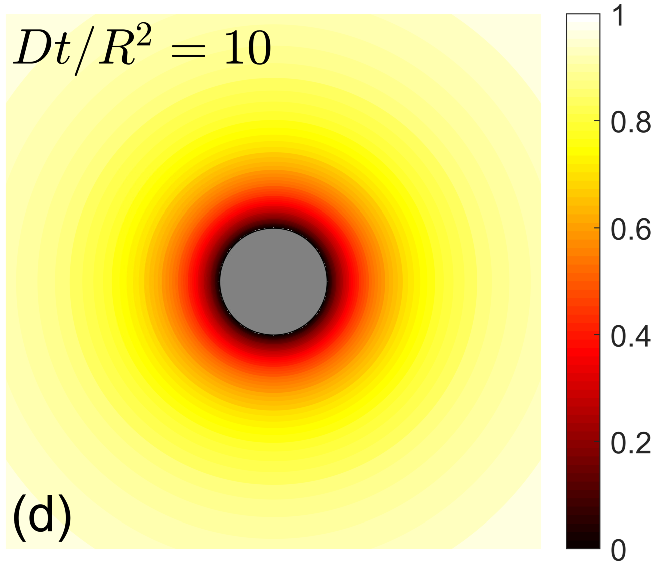}
\includegraphics[width=0.19\textwidth]{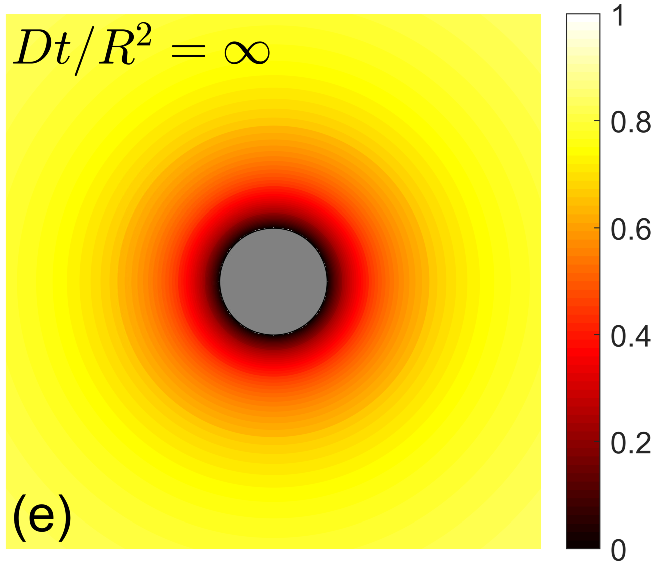}
\includegraphics[width=0.19\textwidth]{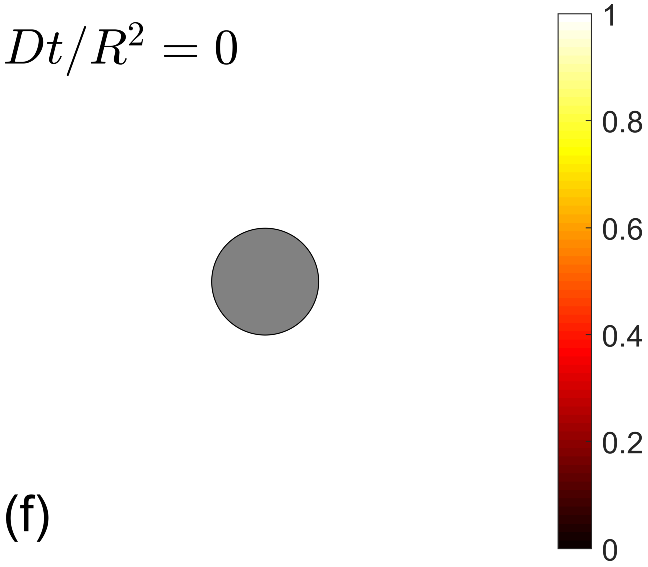}
\includegraphics[width=0.19\textwidth]{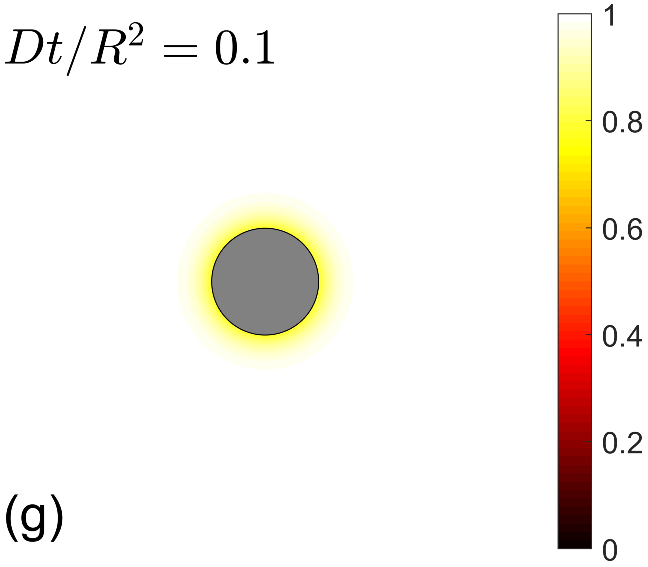}
\includegraphics[width=0.19\textwidth]{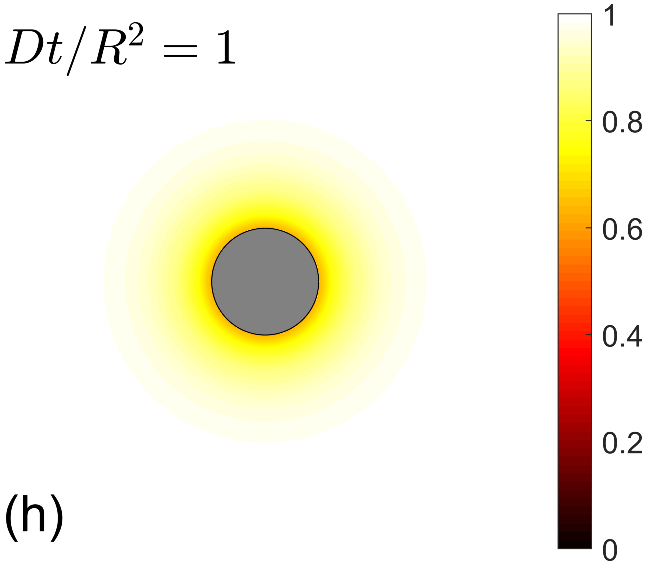}
\includegraphics[width=0.19\textwidth]{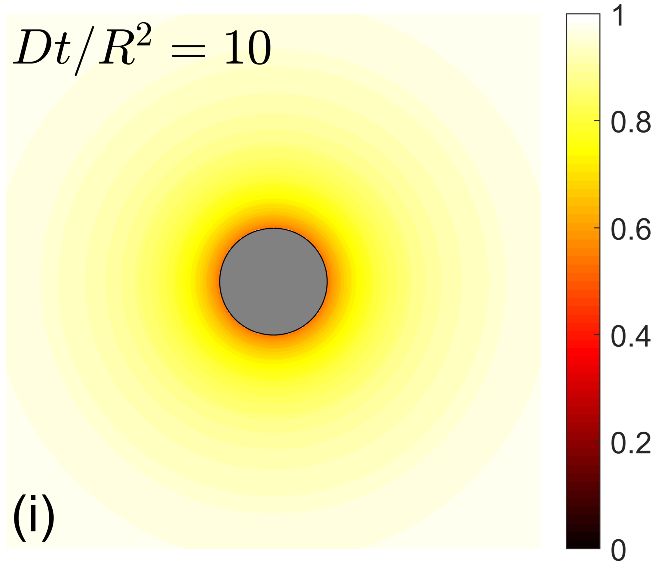}
\includegraphics[width=0.19\textwidth]{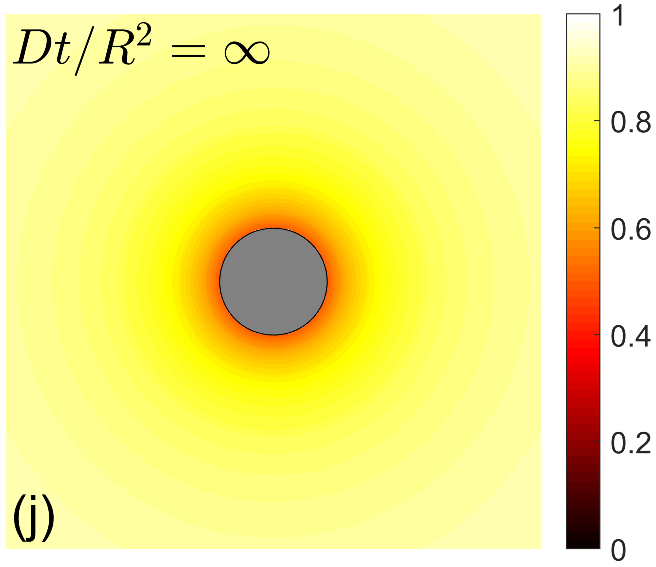}
\caption{
(\textbf{Top row}) Rescaled concentration $[A](\x,t)/[A]_0 = 1 -
\frac{R}{|\x|} \erfc\bigl((|\x|-R)/\sqrt{4D t}\bigr)$ of reactants
$A$ near a perfectly reactive catalytic sphere of radius $R$ (in gray)
at different time instances (here $\erfc(z)$ is the complementary
error function) \cite{Smoluchowski18}.  (\textbf{a}) Homogeneous
concentration at $t = 0$; (\textbf{b}) Formation of a thin depletion
zone at short time $Dt/R^2 = 0.1$; (\textbf{c,d}) Progressive growth
of the depletion zone at larger times $Dt/R^2 = 1$ and $ Dt/R^2 = 10$;
(\textbf{e}) Approach to a steady-state concentration
$[A](\x,\infty)/[A]_0 = 1 - R/|\x|$ as $t\to\infty$.  (\textbf{Bottom
row}) Rescaled concentration $[A](\x,t)/[A]_0 = 1 - \frac{R -
R_\kappa}{|\x|} \left\{ \erfc\left(\frac{|\x|-R}{\sqrt{4Dt}}\right) +
e^{Dt/R_\kappa^2 + (|\x|-R)/R_\kappa}
\erfc\left(\frac{|\x|-R}{\sqrt{4Dt}} +
\frac{\sqrt{Dt}}{R_\kappa}\right)\right\}$ of reactants $A$ near a
partially reactive catalytic sphere of radius $R$, with reactivity
$\kappa R/D = 1$, at the same time instances \cite{Collins49} (with
$R_\kappa = R/(1 + \kappa R/D)$).}
\label{fig:diffusion}
\end{figure}

For the sake of clarity, we focus on heterogeneous catalysis when a
reactant $A$ can be transformed into a product $B$ in the presence of
an immobile catalyst $C$:
\begin{equation}  \label{eq:A+C}
A + C \rightarrow B + C .
\end{equation}
If the catalytic germs were uniformly dispersed in a chemical reactor,
one could still rely on Eq. (\ref{eq:[A]-linear}).  However, in many
practical situations, catalytic germs have specific locations, most
often on a {\it surface} of a porous medium, so that the reactant $A$
should first reach this spatial location.  As the reactants near the
catalytic germs have higher chances to reach on them and thus to be
transformed to $B$, the concentration $[A]$ becomes space-dependent.
In particular, a depletion zone with low concentration of $A$ is
formed near the catalytic surface (Fig. \ref{fig:diffusion}, top row).
This is a direct consequence of the transport step, which can be
described, as in the case of coagulation dynamics, by the diffusion
equation (also called Smoluchowski equation or heat equation),
\begin{equation}  \label{eq:diffusion}
\frac{\partial [A]}{\partial t} = D  \Delta [A],
\end{equation}
where $\Delta = \partial^2/\partial x^2 + \partial^2/\partial y^2 +
\partial^2/\partial z^2$ is the Laplace operator, and $D$ is the
diffusion coefficient of reactant $A$ in a liquid.  In analogy to
Eq. (\ref{eq:[A]-linear}), this equation describes the time evolution
of the concentration $[A](\x,t)$ in each spatial point $\x$ due to
diffusive displacements of the reactants $A$ in the bulk.  In turn,
the reaction itself that occurs on the catalytic surface $C$ is
implemented via a {\it boundary condition} on that surface.  If $A$ is
transformed into $B$ immediately upon the first encounter with the
catalyst $C$ (so-called perfect reactions), the concentration $[A]$ is
set to zero on $C$.  This so-called Dirichlet boundary condition was
first imposed by von Smoluchowski and still remains the most
well-studied and frequently used boundary condition.  Its effect is
illustrated on Fig. \ref{fig:diffusion} (top row) by dark color near
the surface of a spherical catalyst.  Note that the overall reaction
rate is determined by the diffusive flux of reactants $A$ onto the
catalytic surface $C$:
\begin{equation} 
J(t) = \int\limits_C d\x \biggl(-D \frac{\partial [A](\x,t)}{\partial n} \biggr),
\end{equation}
where $\partial/\partial_n = (\vec n \cdot \nabla)$ is the normal
derivative along the normal direction $\vec n$ to the surface.  

The inclusion of space dependence into the theory of chemical kinetics
led to many fundamental changes.  As reaction does not occur
homogeneously in space anymore, there are two consecutive steps: the
diffusion step (transport towards the catalytic surface described by
the diffusion equation) and the reaction step (chemical transformation
from $A$ to $B$ on it described by the boundary condition).  The
dependence of these two steps on the shape of the catalytic surface
introduces a new {\it geometric} dimension to the theory
\cite{Benichou10,Benichou14,Grebenkov19}.  How efficient are catalytic
surfaces of different shapes?  Since reaction occurs on the catalytic
surface, can {\it irregularly shaped} catalysts speed up the overall
production due to their higher surface area?  Can one optimize the
shape to increase the production?  Have the large reactive surfaces of
exchange organs such lungs and placentas been optimized by evolution
for more efficient oxygen capture?  These and many other questions
have been intensively studied since the 80's of the twentieth century
(see
\cite{ben-Avraham,Blender90,Coppens99,Filoche00,Grebenkov05,Andrade07,Serov16,Galanti16,Grebenkov20f} and
references therein).

\section{Imperfect surface reactions}
\label{sec:Robin}

Despite ``popularity'' among theoreticians, perfect surface reactions
ignore intrinsic chemical kinetics during the reaction step and
therefore may lead to paradoxical predictions.  For instance, the
overall reaction rate on a perfectly reactive sphere of radius $R$,
\begin{equation}  \label{eq:J_Smol}
J(t) = 4\pi R D [A]_0 \biggl(1 + \frac{R}{\sqrt{\pi D t}}\biggr), 
\end{equation}
found by von Smoluchowski \cite{Smoluchowski18}, is infinitely large
at the very first time instance (as $t\to 0$).  This divergence is
caused by the molecules in an immediate vicinity of the catalyst that
react instantly.  As a consequence, if one searches to maximize the
overall production by distributing a given amount of a catalytic
material, the optimal solution consists in dispersing this material
into a ``dust'', i.e., a uniform arrangement of tiny catalytic germs.
Moreover, if the subdivision of this material into smaller and smaller
germs could be repeated up to infinity, such a fractal dust would
transform all the reactants $A$ in the bulk instantly \cite{Nguyen10}.
From a mathematical point of view, this is not surprising because any
reactant $A$ would have in its immediate vicinity a tiny catalytic
germ, thus eliminating the diffusion step.  However, such a behavior
does not make sense from a practical point of view.  Limitations of
perfect reactions have been recognized in 1949 by Collins and Kimball
\cite{Collins49} who proposed to replace Dirichlet boundary
condition by so-called Robin or radiative boundary condition on the
catalytic surface:
\begin{equation}  \label{eq:Robin}
-D \frac{\partial [A](\x,t)}{\partial n} = \kappa \, [A](\x,t).
\end{equation}
This condition {\it postulates} that the (net) diffusive flux of
reactants $A$ coming onto the catalytic surface from the bulk (the
left-hand side) is proportional to their concentration $[A]$ on that
surface at each surface point.  The proportionality coefficient
$\kappa$, called a ``reactivity'' of the catalytic surface, can range
from $0$ for an inert surface to infinity for a perfectly reactive
surface. In the former case, the diffusive flux of reactants is zero,
meaning that no reaction occurs on that surface.  In the latter case,
the division by $\kappa$ and the limit $\kappa\to\infty$ reduce
Eq. (\ref{eq:Robin}) back to the Dirichlet boundary condition
$[A](\x,t) =0$ on the surface of $C$.  Note that the reactivity
$\kappa$ (in units m/s) can also be expressed in terms of a forward
reaction constant $k_{\rm on}$ (in units m$^3$/s/mol or 1/M/s) as
$k_{\rm on} = \kappa N_A S_C$, where $N_A$ is the Avogadro number, and
$S_C$ is the surface area of the catalytic surface.  Figure
\ref{fig:diffusion}(bottom row) illustrates the effect of partial
reactivity onto the concentration of reactants near the catalytic
sphere of radius $R$.  The depletion zone is thinner and grows slower
than in the case of perfect reactions.  Moreover, the overall reaction
rate $J(t)$ is finite in the short-time limit: $J(0) = 4\pi R^2 \kappa
[A]_0$.  Indeed, only the molecules near the catalyst (of surface area
$4\pi R^2$) can react at short times, and their contribution is now
limited by chemical kinetics, i.e., by the time needed for chemical
transformation (\ref{eq:A+C}), which is controlled by the reactivity
$\kappa$.  As time increases, molecules from further locations arrive
onto the catalyst and can thus contribute.  At long times, the region
near the catalyst is depleted, and reactants $A$ from very distant
locations need to diffuse towards the catalyst.  In this limit, one
gets $J(\infty) = 4\pi R D [A]_0/(1 + D/(\kappa R))$, i.e., the
overall production is therefore diffusion-limited.  In other words,
the overall production exhibits a transition from the reaction-limited
regime at short times to the diffusion-limited regime at long times.

\begin{figure}[t!]
\includegraphics[width=0.20\textwidth]{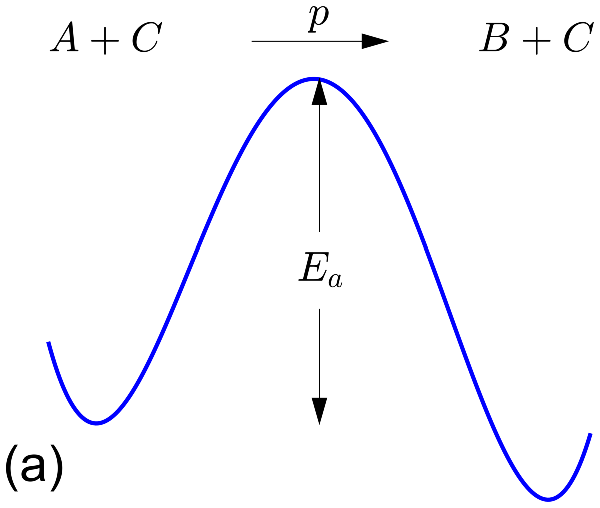}  \hskip 3mm
\includegraphics[width=0.25\textwidth]{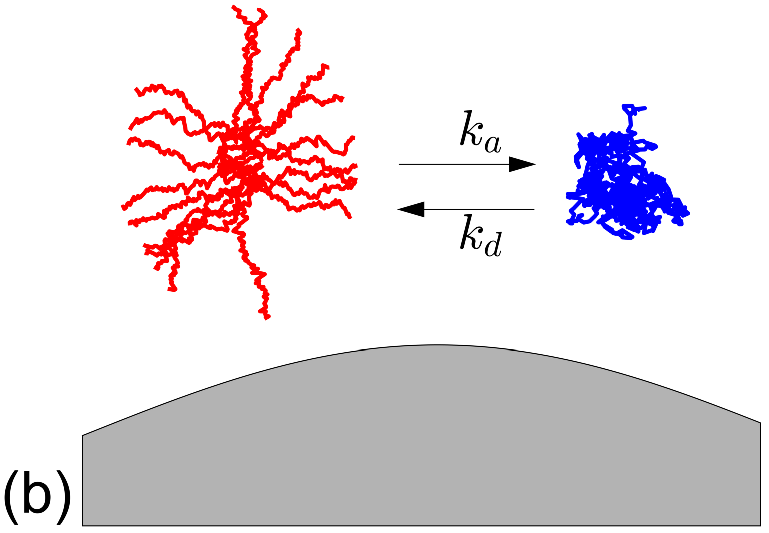}  \hskip 3mm
\includegraphics[width=0.25\textwidth]{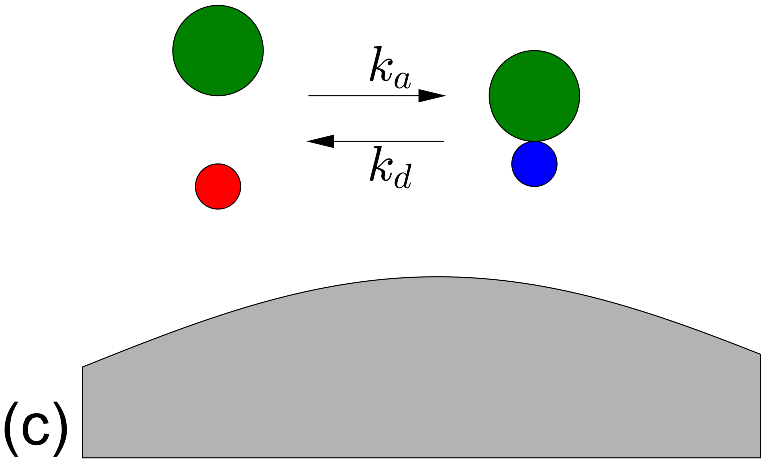}
\includegraphics[width=0.30\textwidth]{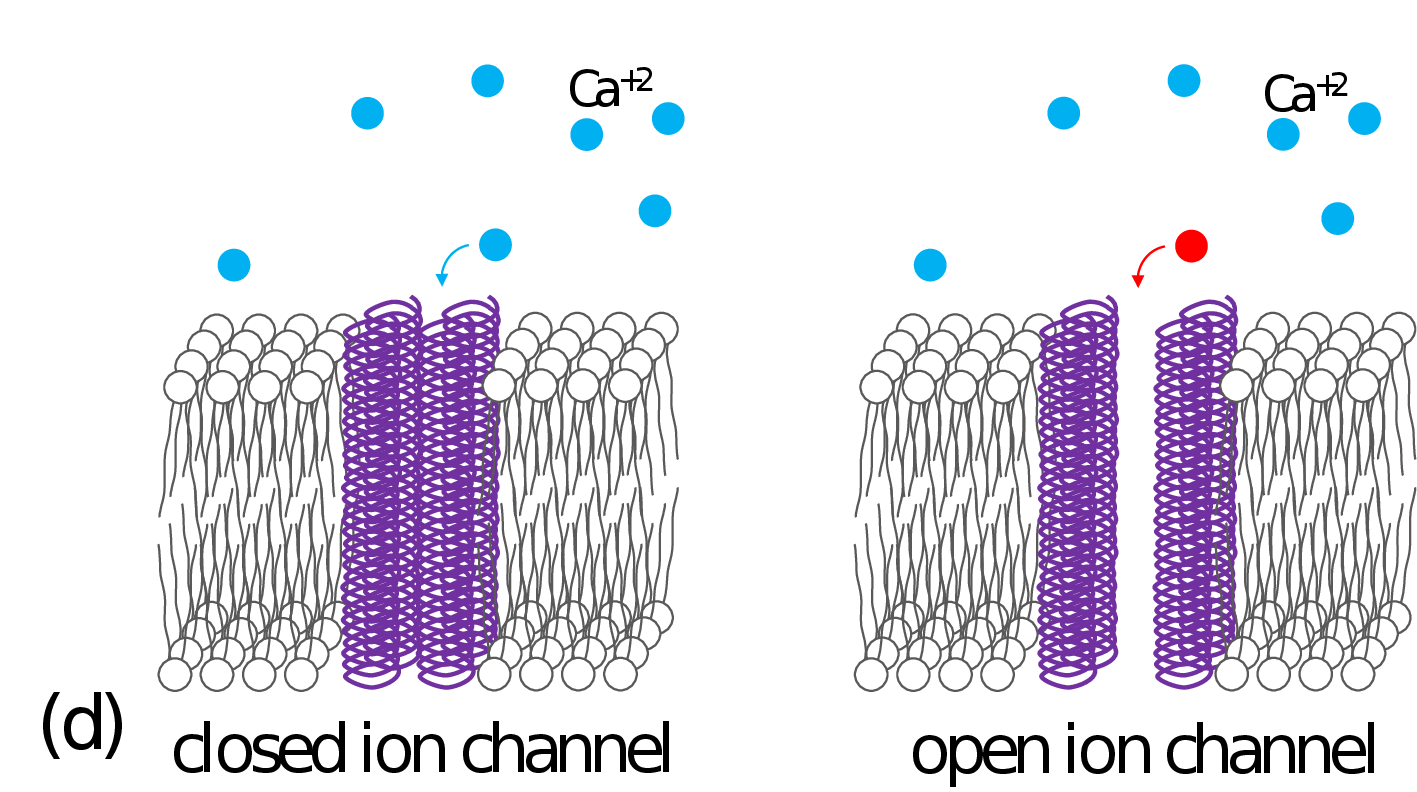}  \hskip 3mm
\includegraphics[width=0.25\textwidth]{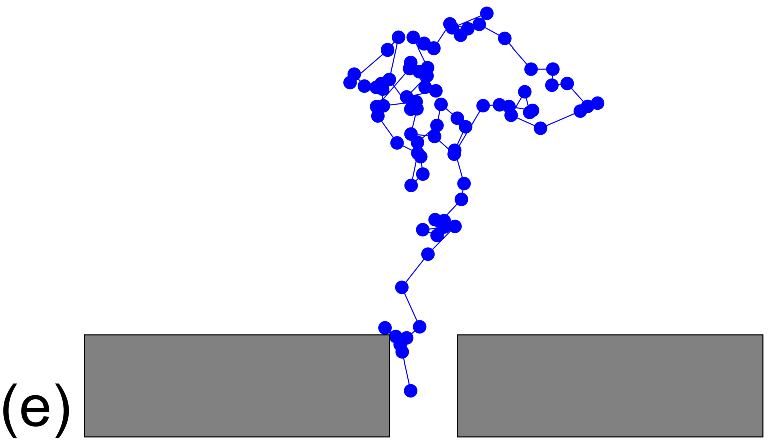}   \hskip 3mm
\includegraphics[width=0.25\textwidth]{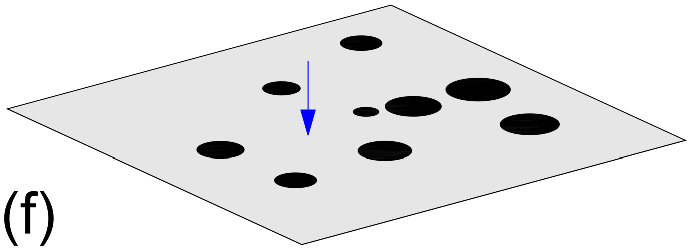}
\caption{
Various microscopic origins of imperfect surface reactions.
(\textbf{a}) When the reactant $A$ arrives onto the catalytic surface
$C$, an activation energy barrier $E_a$ has to be overcome for a
chemical transformation of $A$ into $B$; if failed, the reactant
leaves the vicinity of $C$ and thus resumes its bulk diffusion;
(\textbf{b}) A macromolecule can spontaneously switch its
conformational state from ``active'' (in red) to ``passive'' (in blue)
with the rate $k_a$, and back (with the rate $k_d$), while its
reaction on the catalytic surface (in gray) or with another
macromolecule (a receptor, an enzyme, a DNA strand, etc.) is only
possible in the ``active'' conformational state; (\textbf{c}) The
reactant can be temporarily trapped by a buffer molecule (in green)
that makes it inactive for the considered surface reaction; their
association/dissociation kinetics is usually described by forward and
backward rates $k_a$ and $k_d$; (\textbf{d}) An ion can pass through
an open channel, while it is reflected back from the closed channel;
(\textbf{e}) An escape of a semi-flexible polymer through a small hole
can be described by an entropic barrier that leads to partial
reactivity when the first arrival to the hole does not guarantee the
passage; (\textbf{f}) An inert (gray) surface is covered by reactive
catalytic germs (black spots) so that the reactant may fail to react
upon the first arrival, and thus resumes its bulk diffusion until the
next encounter, and so.  Similarly, a protein can search for a
specific (target) site on a DNA chain for successful binding.}
\label{fig:Robin}
\end{figure}

The partial reactivity of the surface, described by Robin boundary
condition (\ref{eq:Robin}), can model various microscopic mechanisms
of imperfect reactions \cite{Grebenkov19,Piazza22}, as illustrated by
Fig. \ref{fig:Robin}.  In physical chemistry, once the reactant $A$
arrives onto the catalytic surface, it has to overcome an activation
energy barrier in order to react \cite{Weiss86,Hanggi90}.  This
activation energy determines the probability $p$ of the reaction
attempt to be successful.  However, the reactant may fail its reaction
attempt (with probability $1-p$) by leaving the proximity of the
catalytic surface and thus resuming its diffusion until the next
encounter, and so on.  In this setting, the microscopic interaction
determines the probability $p$, which, in turn, fixes the effective
macroscopic reactivity $\kappa = \frac{D}{a} \, \frac{p}{1-p}$, where
$a$ is the width of the reactive layer near the catalytic surface
(i.e., the interaction range, which is typically of the order of a
nanometer) \cite{Grebenkov03}.  Varying $p$ from $0$ to $1$ covers the
whole range of reactivities from $0$ to $+\infty$.  In the biochemical
context, conformational changes of a macromolecule between nearly
isoenergetic folded states can alter its function; this mechanism is
primarily important for protein-ligand and protein-protein recognition
\cite{Cortes10,Luking22,Bressloff17}.  When such a protein arrives
onto the catalytic surface (its reaction partner), it has to be in an
appropriate conformational state (with probability $p$) to be able to
initiate the reaction (\ref{eq:A+C}); otherwise, the protein leaves
the catalytic surface and restarts its bulk diffusion
\cite{Galanti16b}.  Even small particles such as calcium ions can
spontaneously lose their reactivity via reversible binding to buffer
molecules.  This is the basis of one of the regulatory mechanisms in
neuron signaling when tuning the concentration of buffer molecules
inside a presynaptic bouton controls the ability of calcium ions to
reach calcium-sensing proteins that trigger the vesicular release of
neurotransmitters (see \cite{Reva21} and references therein).
In the microcellular context, the catalytic surface may represent a
plasma membrane of a cell or of a nucleus, while the reaction event
may consist in the passage through a channel on that membrane; such
``reaction'' occurs if the channel is open (with probability $p$),
while the reactant is reflected back from a closed channel
\cite{Benichou00,Reingruber09,Lawley15}.  Even if the channel is
always open (e.g., just a hole in a container or in a filter), there
is an entropic barrier that may prohibit the escape from the confining
domain and lead to reflection and resumed diffusion
\cite{Zhou91,Reguera06,Chapman16}.  In heterogeneous catalysis, the
macroscopic reactivity $\kappa$ may account for micro-heterogeneity of
the catalytic surface, which is not fully covered by catalytic germs;
in this case, $p$ is the probability to hit the catalytic surface at
the catalytic germ (and thus to react), while $1-p$ is the probability
of arriving at the inert part of the surface and thus being reflected.
Homogenization of spatially heterogeneous catalytic surfaces leads to
Robin boundary condition (\ref{eq:Robin}), in which the reactivity
$\kappa$ effectively accounts for distributed reactive spots
\cite{Berg77,Berezhkovskii04,Berezhkovskii06,Muratov08,Bernoff18,Punia21}.
For instance, in the seminal work by Berg and Purcell \cite{Berg77},
the probability $p$ was found for a spherical cell of radius $R$
covered by $N$ disk-shaped receptors of radius $a$: $p = Na/(Na + \pi
R)$.

The partial reactivity adds an important intermediate step to
diffusion-controlled reactions: after the first arrival onto the
catalytic surface, the reactant executes a sequence of diffusive
explorations of the bulk near the catalytic surface after each failed
reaction attempt \cite{Grebenkov07b,Erban07,Singer08,Grebenkov20c}.
This step may considerably slow up the overall production, while the
shape and reactivity of the catalytic surface are entangled through
diffusion in a sophisticated way.  Note that the same problem emerges
in the context of semi-permeable membranes in biology and blocking
electrodes in electrochemistry
\cite{Sapoval94,Sapoval96,Grebenkov06}.  The role of reactivity (or,
equivalently, permeability or resistivity) onto the overall production
was thoroughly investigated
\cite{Sano79,Brownstein79,Powles92,Bressloff08,Grebenkov17a,Guerin21}.
For instance, B. Sapoval and co-workers discussed the role of the
``reaction length'' $D/\kappa$ as a physical scale for oxygen capture
efficiency in human lungs \cite{Sapoval02}.

\section{Various extensions}
\label{sec:extension}

The basic description of diffusion-controlled reactions via
Eqs. (\ref{eq:diffusion}, \ref{eq:Robin}) has been generalized in
different ways.  Most efforts were dedicated to extensions of the
diffusion equation (\ref{eq:diffusion}) that describes the simplest
diffusive motion of reactants, the so-called Brownian motion.  For
instance, the Fokker-Planck equation allows one to incorporate the
effects of external potentials (e.g., an electric field acting on a
charged particle), anisotropy, and space- and/or time-dependence of
the diffusion coefficient \cite{Gardiner,Risken,VanKampen,Schuss}.
Fractional space and time derivatives can further include nonlocal
displacements and memory effects in continuous-time random walks
\cite{Metzler00,Metzler04,Sokolov12,Krapf19}.  Diffusing diffusivity
and switching diffusivity models were proposed to describe the
diffusive transport in dynamically heterogeneous media or in the
presence of buffer molecules that may reversibly bind the reactant and
thus randomly change its diffusion coefficient
\cite{Chubynsky14,Chechkin17,Lanoiselee18,Grebenkov19d}.  The addition
of a linear term proportional to $[A]$ to the right-hand side of the
diffusion equation (\ref{eq:diffusion}) can account for first-order
disintegration mechanisms such as photo-bleaching, bulk relaxation,
radioactive decay, or a finite lifetime of the reactant
\cite{Yuste13,Meerson15,Grebenkov17}, as well as the effect of 
diffusion-sensitizing magnetic field gradient encoding in diffusion
magnetic resonance imaging \cite{Grebenkov07}.  Moreover, the
diffusion equation with nonlinear terms in $[A]$ can describe reaction
waves and many out-of-equilibrium chemical reactions involving
``activators'' and ``inhibitors'' (e.g., Belousov-Zhabotinsky
reaction), paving a way to the theory of pattern formations initiated
by A. Turing \cite{Murrey,Turing52}.

The above extensions generally employ the canonical Dirichlet or Robin
boundary conditions.  Such a ``persistence'' can partly be explained
by two mathematical reasons: (i) the Laplace operator with either of
these boundary conditions is known to be self-adjoint (Hermitian) that
allows one to rely on powerful methods of spectral theory and to
borrow numerous tools from quantum mechanics; (ii) the diffusion
equation with these boundary conditions has a straightforward
probabilistic interpretation that provides strong intuition onto the
behavior of the studied diffusion-reaction processes, offers efficient
Monte Carlo simulations, and helps to extend the macroscopic
description in terms of concentrations to single-molecule experiments.
In fact, many biochemical reactions involve proteins that are not
abundant inside living cells.  When the number of proteins is
relatively small (e.g., few tens or few hundred of transcription
factors in a bacterium \cite{Milo}), the macroscopic notion of
concentration may be inapplicable, the overall reaction rate may be
uninformative or even misleading, while {\it fluctuations} become
critically important.  Such reactions require therefore a
probabilistic description in terms of the survival probability of a
single reactant molecule and the probability density of the
first-reaction time \cite{Redner,Metzler,Masoliver}.  In many
settings, the survival probability of a single molecule obeys the same
equations (\ref{eq:diffusion}, \ref{eq:Robin}) and hence is equal to
the rescaled concentration $[A](\x,t)/[A]_0$.  This equivalence
bridges the macroscopic and probabilistic descriptions, providing
complementary insights and opening efficient ways to analyze and
interpret single-molecule experiments
\cite{Yu06,Raj08,Xie08,Li11,Kastantin12,Wang17,Norregaard17,Sungkaworn17,Grebenkov18c,Grebenkov18d,Elf19}.

At the same time, the Robin boundary condition (\ref{eq:Robin})
remains limited to modeling rather simple surface reactions with a
{\it constant} reactivity.  Consideration of time- and/or
space-dependent reactivity is one natural extension (see
\cite{Grebenkov19c} and references therein).  Another important
extension concerns reversible reactions such as binding/unbinding,
association/dissociation, and adsorption/desorption kinetics, in which
case the reactant can be temporarily bound to the surface (or to
another molecule).  The exchange between free particles and those
bound on the surface can be incorporated through the ``back-reaction''
boundary condition, also known as ``generalized radiation'' or
``generalized Collins-Kimball'' boundary condition
\cite{Goodrich54,Mysels82,Agmon84,Agmon89,Agmon90,Kim99,Prustel13,Scher23,Grebenkov23b}.
Application of the Laplace transform with respect to time,
$\tilde{[A]}(\x,s) = \int\nolimits_0^\infty dt \, e^{-st} \,
[A](\x,t)$, reduces this boundary condition to Robin boundary
condition (\ref{eq:Robin}) with $s$-dependent reactivity $\kappa(s)$
(see details in \cite{Grebenkov23b}).  In this way, reversible and
irreversible diffusion-controlled reactions admit essentially the same
mathematical description in Laplace domain (in terms of $s$); in turn,
the $s$-dependent reactivity results in fundamentally different
behaviors in time domain (in terms of $t$).  In addition, one can
further relax the assumption of an immobile bound state and allow for
diffusion on the surface.  The efficiency of such intermittent search
dynamics with alternating phases of bulk and surface diffusion was
thoroughly investigated
\cite{Chechkin09,Benichou10b,Rojo11,Chechkin11,Chechkin12,Berezhkovskii15,Berezhkovskii17}
(see also a review \cite{Benichou11}).

\section{Beyond the conventional framework}
\label{sec:encounter}

To handle more general surface reaction mechanisms such as, e.g.,
deactivation or passivation of catalysts \cite{Filoche05,Filoche08},
or progressive activation of enzymes, an alternative theoretical
description of diffusion-controlled reactions was proposed
\cite{Grebenkov20}.  This so-called encounter-based approach
originates from the theory of reflected stochastic processes in
confined domains and relies on the concept of the boundary local time
$\ell$ -- a rescaled number of encounters between the reactant and the
catalytic surface.  In this approach, one can {\it disentangle} the
respective roles of the shape and reactivity of the catalytic surface.
In fact, the concentration of reactants $A$ can be represented as
\begin{equation}  \label{eq:A_rho}
[A](\x,t) = \int\limits_0^\infty d\ell \, e^{-\ell \kappa/D} \, \rho(\ell,\x,t),
\end{equation}
where $\rho(\ell,\x,t)$ describes the statistics of encounters with an
inert surface.  In other words, the function $\rho(\ell,\x,t)$ encodes
how the shape of the catalytic surface affects the diffusive dynamics,
whereas the exponential factor $e^{-\ell \kappa/D}$ incorporates the
reactivity $\kappa$ that was {\it implicitly} imposed via Robin
boundary condition (\ref{eq:Robin}) in the conventional approach.  As
the successful surface reaction is preceded by a sequence of failed
reaction attempts at each encounter, the exponential factor in
Eq. (\ref{eq:A_rho}) can be interpreted as the exponential probability
law, $\P\{a \hat{n} > \ell\} = e^{-\ell \kappa/D}$, for the random
number $\hat{n}$ of encounters in that sequence.  Due to the
self-similar nature of Brownian motion, the number of encounters has
to be rescaled by the width $a$ of a thin surface layer, in which the
molecule can interact with the catalytic surface (see details in
\cite{Grebenkov20}).  While the statistics of encounters was
investigated for simple confinements
\cite{Grebenkov07b,Grebenkov19b,Grebenkov20c,Bressloff22,Grebenkov22},
its shape dependence for porous media representing industrial
catalysts or biological environments remains still unknown.

\begin{figure}[t!]
\includegraphics[width=80mm]{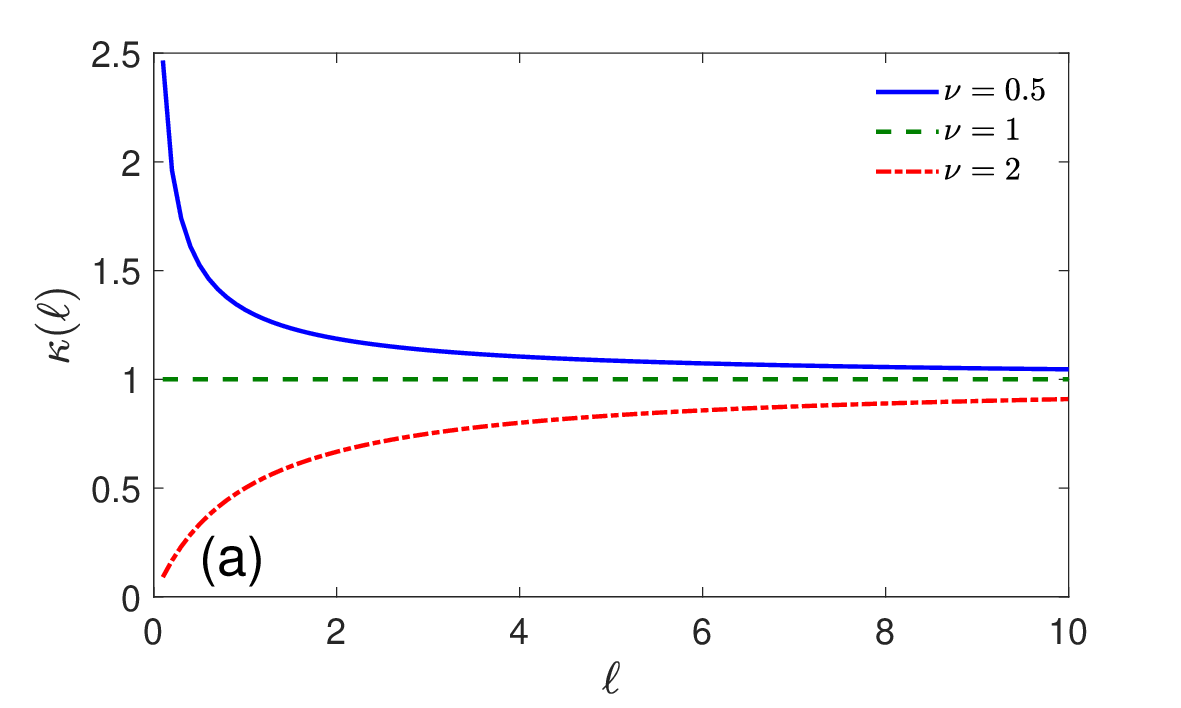}
\includegraphics[width=80mm]{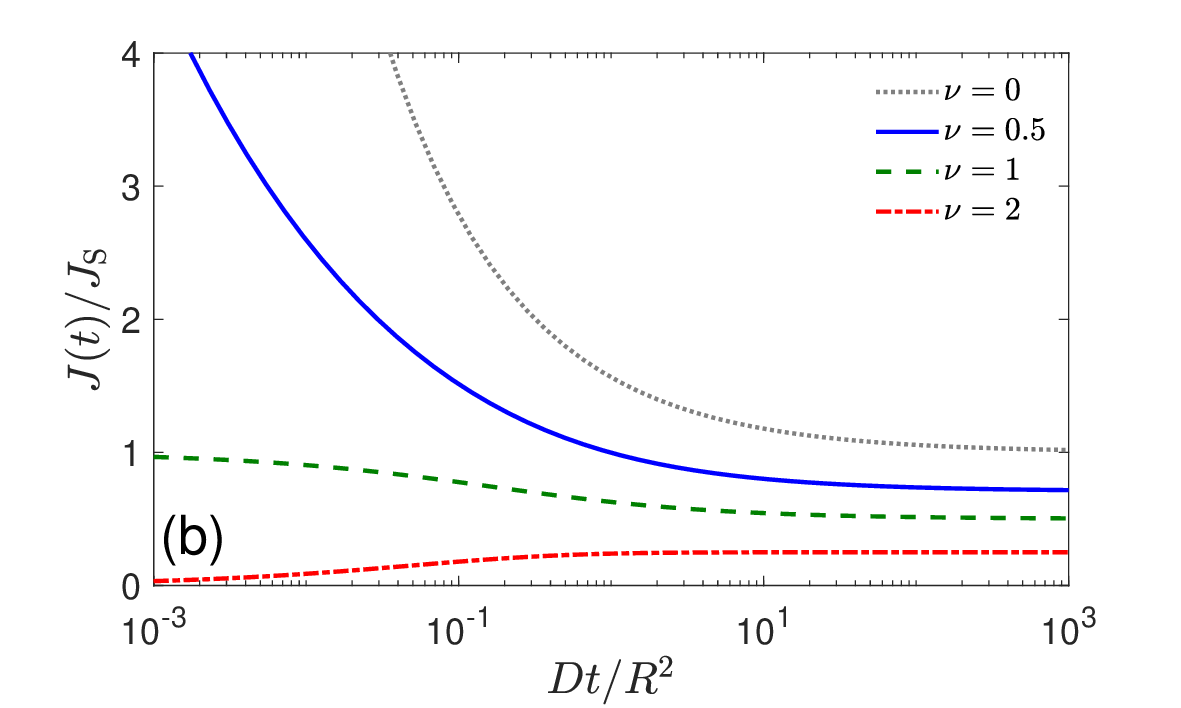}
\caption{
(\textbf{a}) Encounter-dependent reactivity $\kappa(\ell)$ from the
gamma model, with $q=1$ and three values of $\nu$.  (\textbf{b}) The
overall reaction rate $J(t)$ on a spherical catalyst of radius $R$,
rescaled by the Smoluchowski's rate $J_S=4\pi DR[A]_0$, with $q=1$ and
three values of $\nu$.  Dotted curve represents Eq. (\ref{eq:J_Smol})
for a perfectly reactive sphere (it formally corresponds to $\nu
=0$).}
\label{fig:flux}
\end{figure}

Most importantly, one can replace the exponential factor in
Eq. (\ref{eq:A_rho}), which incorporated the effect of a constant
reactivity $\kappa$, by another probability law $\P\{a \hat{n} >
\ell\} = \Psi(\ell)$, to model more sophisticated surface reaction
mechanisms with an encounter-dependent reactivity
\begin{equation}   \label{eq:kappa}
\kappa(\ell)= D \frac{-\frac{d}{d\ell} \Psi(\ell)}{\Psi(\ell)} \,.
\end{equation}
If $\Psi(\ell) = e^{-\ell \kappa/D}$, this formula yields the constant
reactivity considered above, $\kappa(\ell)= \kappa$, and ensures the
Markovian character of the binding reaction.  However, another {\it
choice} of the function $\Psi(\ell)$ allows one to implement the
reactivity of the catalytic surface that depends on how many times the
reactant has encountered it.  To illustrate this idea, let us consider
the gamma model, by choosing $\Psi(\ell)=
\Gamma(\nu,q\ell)/\Gamma(\nu,0)$, where $q > 0$ and $\nu > 0$ are two
parameters, and $\Gamma(\nu,z) =
\int\nolimits_z^\infty dx \, x^{\nu-1} e^{-x}$ is the upper incomplete
gamma function.  For $\nu = 1$, one has $\Gamma(1,z) = e^{-z}$ and
thus retrieves the above setting of constant reactivity $\kappa = qD$.
Figure \ref{fig:flux} illustrates the corresponding
encounter-dependent reactivity $\kappa(\ell)$, given by
Eq. (\ref{eq:kappa}) (panel a), and the overall reaction rate $J(t)$
on a spherical catalyst of radius $R$ (panel b) that can be found in
the framework of the encounter-based approach \cite{Grebenkov20}.
When $0< \nu <1$, the catalytic surface is highly reactive at the
beginning and then reaches a constant reactivity $q D$.  This
situation can model a progressive passivation of the catalytic surface
by repeated encounters with the reactant, up to a constant level.
Expectedly, the diffusive flux is high at short times and then
decreases to a constant steady-state level.  Note that $\nu =0$
formally corresponds to a perfect reaction, with the Smoluchowski's
rate (\ref{eq:J_Smol}).  The particular value $\nu =1$ yields the
constant reactivity, independent of the number of encounters, for
which the diffusive flux is constant at short times, $4\pi R^2
qD[A]_0$, and slowly decreases to another constant at long times, as
predicted by Collins and Kimball \cite{Collins49}.  In turn, if $\nu
>1$, the catalytic surface is passive at the beginning and then
reaches a constant reactivity.  This situation can model progressive
activation of that catalytic surface.  Accordingly, the overall
reaction rate is zero at short times and then increases to a constant
steady-state level.  Choosing an appropriate function $\Psi(\ell)$,
one can produce the desired shape of the encounter-dependent
reactivity $\kappa(\ell)$ that opens a way to model various surface
reaction mechanisms.

The encounter-based approach goes far beyond the conventional theory
of diffusion-controlled reactions described by Dirichlet or Robin
boundary conditions.  From the mathematical point of view, the
description of a general surface reaction with an encounter-dependent
reactivity $\kappa(\ell)$ is not reducible to the Robin boundary
condition.  As a consequence, Laplacian eigenfunctions that are
conventionally used in spectral expansions, need to be replaced by
so-called Steklov eigenfunctions \cite{Grebenkov20,Levitin}.  Though
being less known in the context of chemical reactions, these
eigenfunctions turn out to be particularly well suited for describing
diffusive explorations near a catalytic surface between successive
reaction attempts.  Several extensions of the encounter-based approach
have already been explored such as (i) inclusion of an external
potential that leads to a biased or drifted motion
\cite{Grebenkov22a}; (ii) the effects of stochastic resetting
\cite{Evans11,Evans20} of the position and of the boundary local time
onto diffusion-controlled reactions \cite{Bressloff22d,Benkhadaj22};
(iii) the cooperative search by multiple independent particles and the
related extreme first-passage statistics \cite{Grebenkov22b}; (iv) the
escape problem \cite{Grebenkov23a}; (v) non-Markovian
binding/unbinding kinetics \cite{Grebenkov23b}.  Moreover, the same
concepts can be applied to describe diffusive permeation across
membranes \cite{Bressloff22c,Bressloff23a,Bressloff23b}.  Despite
these recent advances, there are many open questions and promising
perspectives for future developments, such as merging anomalous bulk
diffusions with generalized surface reactions, the effect of
sophisticated geometric confinements onto the encounter statistics,
competition of multiple reactive centers for capturing a limited
amount of diffusing reactants, indirect coupling of different
reactants through encounter-dependent catalytic surfaces, inference of
appropriate surface reaction models from experimental data, to name
but a few.

\section{Conclusion}
\label{sec:conclusion}

In summary, we reviewed the major steps in the long history of
developments in the theory of diffusion-controlled reactions.  M. von
Smoluchowski first recognized the importance of the diffusion step,
during which the reactants have to meet each other.  He also put
forward the diffusion equation to describe of the dynamics of
reactants in the bulk and boundary conditions to account for the
reaction on the surface.  His mechanism of perfect reactions upon the
first encounter was then improved by Collins and Kimball.  While most
later theoretical efforts were dedicated to improvements of the bulk
dynamics, an encounter-based approach was recently developed to enable
more general surface reaction mechanisms.  This approach has already
shown many advantages such as probabilistic insights onto surface
reactions, disentanglement of the impacts of shape and reactivity of
the catalytic surface, flexibility in characterization of diffusive
explorations near the reactive surface, etc.  In particular, the
concept of encounter-dependent reactivity allows one to describe an
action of reactants onto the catalytic surface, and such a feedback
may potentially be relevant in various biochemical and electrochemical
settings.  There are still many open questions and current
developments, aiming at understanding the mathematical formalism of
the encounter-based approach, relating the shape of the catalytic
surface to the spectral properties of the underlying operators,
elaborating various extensions, and uncovering potential applications
in chemistry and biochemistry.

\vspace{6pt} 

\begin{acknowledgments}
The author acknowledges the Alexander von Humboldt Foundation for
support within a Bessel Prize award.
\end{acknowledgments}


\end{document}